
%
\input harvmac

\def\Titlehhh#1#2{\nopagenumbers\abstractfont\hsize=\hstitle\rightline{#1}%
\vskip .1in\centerline{\titlefont #2}\abstractfont\vskip .1in\pageno=0}
\def\abstractfont{\ninepoint}
\def\CTPa{\it Center for Theoretical Physics, Department of Physics,
      Texas A\&M University}
\def\CTPb{\it College Station, TX 77843-4242, USA}
\def\HARCa{\it Astroparticle Physics Group,
Houston Advanced Research Center (HARC)}
\def\HARCb{\it The Woodlands, TX 77381, USA}

\def\ie{\hbox{\it i.e.}}     
\def\eg{\hbox{\it e.g.}}     

\def\nextline{\unskip\nobreak\hfill\break}

\catcode`\@=11 

\def\lsim{\mathrel{\mathpalette\@versim<}}
\def\gsim{\mathrel{\mathpalette\@versim>}}
\def\@versim#1#2{\vcenter{\offinterlineskip
    \ialign{$\m@th#1\hfil##\hfil$\crcr#2\crcr\sim\crcr } }}
\def\boxit#1{\vbox{\hrule\hbox{\vrule\kern3pt
      \vbox{\kern3pt#1\kern3pt}\kern3pt\vrule}\hrule}}

\def\etal{{\it et. al.}}
\def\r#1{$\bf#1$}
\def\rb#1{$\bf\overline{#1}$}

\def\t1{{\tilde 1}}

\def\JL{J. L. Lopez}
\def\DVN{D. V. Nanopoulos}
\def\slash#1{#1\hskip-6pt/\hskip6pt}

\def\GeV{\,{\rm GeV}}
\def\TeV{\,{\rm TeV}}

\def\pb{\,{\rm pb}}
\def\fb{\,{\rm fb}}
\def\ipb{\,{\rm pb^{-1}}}

\def\NPB#1#2#3{Nucl. Phys. B {\bf#1} (19#2) #3}
\def\PLB#1#2#3{Phys. Lett. B {\bf#1} (19#2) #3}

\def\PRD#1#2#3{Phys. Rev. D {\bf#1} (19#2) #3}
\def\PRL#1#2#3{Phys. Rev. Lett. {\bf#1} (19#2) #3}
\def\PRT#1#2#3{Phys. Rep. {\bf#1} (19#2) #3}
\def\MODA#1#2#3{Mod. Phys. Lett. A {\bf#1} (19#2) #3}

\def\TAMU#1{Texas A \& M University preprint CTP-TAMU-#1}

\nref\EKN{J. Ellis, S. Kelley and D. V.  Nanopoulos, \PLB{249}{90}{441},
\PLB{260}{91}{131}; P. Langacker and M.-X. Luo, \PRD{44}{91}{817};
F. Anselmo, L. Cifarelli, A. Peterman, and A. Zichichi, Nuovo Cim. {\bf104A}
(1991) 1817.}
\nref\MATS{M. Matsumoto, J. Arafune, H. Tanaka, and K. Shiraishi,
\PRD{46}{92}{3966}.}
\nref\ANabc{R. Arnowitt and P. Nath, \PRL{69}{92}{725}; P. Nath and
R. Arnowitt, \PLB{287}{92}{89} and \PLB{289}{92}{368}.}
\nref\LNP{\JL, \DVN, and H. Pois, \TAMU{61/92} and CERN-TH.6628/92 (to appear
in Phys. Rev. D).}
\nref\HMY{J. Hisano, H. Murayama, and T. Yanagida, \PRL{69}{92}{1014} and
Tohoku University preprint TU--400 (July 1992).}
\nref\LNPZ{\JL, \DVN, H. Pois, and A. Zichichi, \PLB{299}{93}{262}.}
\nref\LNZa{\JL, \DVN, and A. Zichichi, \PLB{291}{92}{255}.}
\nref\ANdm{R. Arnowitt and P. Nath, \TAMU{65,66/92} and NUB-TH-3055,3056/92.}
\nref\LNZb{\JL, \DVN, and A. Zichichi, \TAMU{68/92}, CERN-TH.6667/92, and
CERN-PPE/92-188.}
\nref\LNWZ{\JL, \DVN, X. Wang, and A. Zichichi, \TAMU{76/92}, CERN/LAA/92-023,
and CERN-PPE/92-194.}
\nref\LNPWZh{\JL, \DVN, H. Pois, X. Wang and A. Zichichi, \TAMU{05/93}.}
\nref\BFM{A. Bartl, H. Fraas, and W. Majerotto, Z. Phys. C {\bf30} (1986) 441.}
\nref\Dionisi{C. Dionisi, \etal, in Proceedings of the ECFA Workshop on LEP
200, Aachen, 1986, ed. by A. B\"ohm and W. Hoogland, p. 380.}
\nref\Tata{M. Chen, C. Dionisi, M. Martinez, and X. Tata, \PRT{159}{88}0{201}.}
\nref\Felcini{M. Felcini in {\it Ten years of susy confronting experiment},
CERN-TH.6707/92--PPE/92-180 (November 1992).}
\nref\GrivazI{J.-F. Grivaz, LAL preprint 92-64.}
\nref\Kat{S. Katsanevas, talk given at the 1993 Aspen Winter Conference.}
\nref\BFMsf{A. Bartl, H. Fraas, and W. Majerotto, Z. Phys. C {\bf34} (1987)
411.}
\nref\HHG{J. Gunion, H. Haber, G. Kane, and S. Dawson, {\it The Higgs Hunter's
Guide}, Addison-Wesley, Redwood City 1990.}
\nref\ERZ{Y. Okada, M. Yamaguchi, and T. Yanagida, Prog. Theor. Phys.
{\bf85} (1991) 1 and \PLB{262}{91}{54}; J. Ellis, G. Ridolfi, and F. Zwirner,
\PLB{257}{91}{83}; H. Haber and R. Hempfling, \PRL{66}{91}{1815}.}
\nref\aspects{S. Kelley, \JL, \DVN, H. Pois, and K. Yuan, \TAMU{16/92} and
CERN-TH.6498/92 (to appear in Nucl. Phys. B).}
\nref\Wu{S. L . Wu, \etal, in Proceedings of the ECFA Workshop on LEP 200,
Aachen, 1986, ed. by A. B\"ohm and W. Hoogland, p. 312.}
\nref\KLNPYh{S. Kelley, \JL, \DVN, H. Pois, and K. Yuan, \PLB{285}{92}{61}.}
\nref\QCD{E. Braaten and J. P. Leveille, \PRD{22}{80}{715};
M. Drees and K. Hikasa, \PLB{240}{90}{455}; S. G. Gorishny, \etal,
\MODA{5}{90}{2703}.}
\nref\Barger{V. Barger, M. S. Berger, A. L. Stange and R. J. N. Phillips,
\PRD{45}{92}{4128}.}
\nref\Weiler{T. J. Weiler and T. C. Yuan, \NPB{318}{89}{337}.}
\nref\Kunszt{Z. Kunszt and F. Zwirner, \NPB{385}{92}{3}.}
\nref\Treille{D. Treille in {\it Ten years of susy confronting experiment},
CERN-TH.6707/92--PPE/92-180 (November 1992).}
\nref\BargerII{V. Barger, K. Cheung, R. J. N. Phillips, and A. L. Stange,
\PRD{46}{92}{4914}.}
\nref\GrivazII{J.-F. Grivaz in {\it Ten years of susy confronting experiment},
CERN-TH.6707/92--PPE/92-180 (November 1992).}
\nref\LNPWZd{\JL, \DVN, H. Pois, X. Wang and A. Zichichi, in preparation.}
\nref\ENO{J. Ellis, \DVN, and K. Olive, \TAMU{75/92} and CERN-TH.6721/92.}

\nfig\I{The cross section for $e^+e^-\to\chi^+_1\chi^-_1$ at $\sqrt{s}=200\GeV$
as a function of the chargino mass ($m_{\chi^\pm_1}$) for the minimal $SU(5)$
supergravity model (top row) and the no-scale flipped $SU(5)$ supergravity
model (bottom row). The smaller size of the latter is due to destructive
interference effects in the presence of a light sneutrino.}
\nfig\II{The leptonic and hadronic branching ratios for the chargino in the
flipped model and  $m_t=100\GeV$. The former ranges from $\approx2/3$
when the slepton-exchange channels dominate, down to $\approx2/9$ when the
$W$-exchange channels dominate, and through very small values when the two
channels interfere destructively. A complementary effect is seen to happen
for the hadronic branching ratio.}
\nfig\III{The number of `mixed' (1-lepton+2-jets+$\slash{p}$) events per ${\cal
L}=100\ipb$ for both models. Note that the predictions on the top row {\it do
not overlap} with those on the bottom one, and therefore if
$m_{\chi^\pm_1}<100\GeV$, LEPII
should be able to exclude at least one of the models.}
\nfig\IV{The number of dilepton events per ${\cal L}=100\ipb$ to be expected
from the process $e^+e^-\to\chi^+_1\chi^-_1$ for the flipped model. The
corresponding number in the minimal $SU(5)$ model is $1/6$ of that shown on the
top row in Fig. 3.}
\nfig\V{The cross section for $e^+e^-\to\chi^0_1\chi^0_2$ for the flipped
model as a function of the chargino mass. Note that chargino masses up to
$\approx130\GeV$ could be explored with this process. Also shown is the
number of dilepton events per ${\cal L}=500\ipb$.}
\nfig\XII{The number of di-electron events per ${\cal L}=100\ipb$ from
selectron
pair production (top row) and the corresponding number of di-taus from
stau pair production (bottom row), as a function of the lightest charged
slepton mass of the corresponding flavor. The results for the di-muons from
smuon pair production are very similar to the number of di-taus.}
\nfig\XIII{The number of di-electrons from selectron pair production as in
\XII, but plotted against the chargino mass (top row), showing that one could
indirectly probe chargino masses as high as 150 GeV. Also shown (bottom row)
are the chargino dileptons from \IV\ which is comprised of 25\%,25\%,50\%\
$ee,\mu\mu,e\mu$ dileptons respectively.}
\nfig\VI{The cross section at $\sqrt{s}=200\GeV$
for $e^+e^-\to Z^*h\to h\nu \bar \nu$ for the minimal $SU(5)$ model (top row)
and the flipped model (bottom row) as a function of the Higgs mass $m_h$. In
the minimal $SU(5)$ model the cross section differs negligibly from the SM
result, whereas in the flipped model decreases of up to $\approx2/3$ are
possible for a small ($<1\%$) set of points in the allowed parameter space
corresponding to relatively light values of $m_A$.  Note the `tail' in the
cross section in the flipped case for $m_h\gsim 110\GeV$ when the second $Z$
goes off-shell.}
\nfig\VII{The branching ratio for $h\rightarrow b\bar b$ as a function of
the Higgs mass $m_h$ for the minimal $SU(5)$ model (top row), and the flipped
model (bottom row). Note the significant departures from the SM result
($\approx85\%$) due to the opening of the $h\to\chi^0_1\chi^0_1$ (and to a
lesser extent $h\to gg$) channel.}
\nfig\VIII{The branching ratio for $h\rightarrow \chi^0_1\chi^0_1$
as a function of the Higgs mass $m_h$ for the minimal $SU(5)$ model (top row)
and the flipped model (bottom row).}
\nfig\IX{The branching ratio for $h\rightarrow gg$ as a function of the Higgs
mass $m_h$ for the minimal $SU(5)$ model (top row) and the flipped
model (bottom row). For the minimal $SU(5)$ model note the few points where the
branching ratio can be quite large, corresponding to a very light $\tilde
t_1$.}
\nfig\X{The branching ratio for $h\rightarrow c\bar c$ as a function of the
Higgs mass $m_h$ for the minimal $SU(5)$ model (top row), and the flipped
model (bottom row).}
\nfig\XI{The number of mixed chargino events as shown in Fig. 3 but versus the
lightest Higgs boson mass instead. All points with $m_h<60\GeV$ are actually
experimentally excluded.}

\centerline{EUROPEAN ORGANIZATION FOR NUCLEAR RESEARCH}
\medskip
\Titlehhh{\vbox{\baselineskip12pt
\hbox{CERN-PPE/93--16}
\hbox{8 February, 1993}
\hbox{CERN-TH.6773/93}
\hbox{CERN/LAA/93--01}
\hbox{CTP--TAMU--89/92}\hbox{ACT--25/92}}}
{\vbox{\centerline{Sparticle and Higgs Production and Detection}
\centerline{at LEPII in two Supergravity Models}}}
\centerline{JORGE~L.~LOPEZ$^{(a)(b)}$, D.~V.~NANOPOULOS$^{(a)(b)(c)}$,
H. POIS$^{(a)(b)}$,}
\centerline{XU WANG$^{(a)(b)}$, and A. ZICHICHI$^{(d)}$}
\centerline{$^{(a)}$\CTPa}
\centerline{\CTPb}
\centerline{$^{(b)}$\HARCa}
\centerline{\HARCb}
\centerline{$^{(c)}${\it CERN Theory Division, 1211 Geneva 23, Switzerland}}
\centerline{$^{(d)}${\it CERN, Geneva, Switzerland}}
\centerline{ABSTRACT}
We study the most promising signals for supersymmetry at LEPII in the context
of two well motivated supergravity models: (i) the minimal $SU(5)$ supergravity
model including the stringent constraints from proton stability and a not too
young Universe, and (ii) a recently proposed string-inspired no-scale flipped
$SU(5)$ supergravity model. Our computations span the neutralino, chargino,
slepton, and Higgs sectors together with their interconnections in this class
of models. We find that the number of `mixed'
(1-lepton + 2-jets + $\slash{p}$) events occuring in the decay of pair-produced
charginos ($\chi^\pm_1$) is quite significant (per ${\cal L}=100\ipb$) for both
models and that these predictions do not overlap. That is, if
$m_{\chi^\pm_1}<100\GeV$ then LEPII should be able to exclude at least one
of the two models. In the no-scale flipped $SU(5)$ model we find that the
number of acoplanar di-electron events from selectron pair production should
allow for exploration of selectron masses up to the kinematical limit and
chargino masses indirectly as high as 150 GeV. We find that the cross section
$e^+e^-\to Z^*h$ deviates negligibly from the SM result in the minimal model,
whereas it can be as much as $1/3$ lower in the flipped model. The usually
neglected invisible mode $h\to\chi^0_1\chi^0_1$ can erode the preferred $h\to
2\,{\rm jets}$ signal by as much as $40\%$ in these models. We conclude that
the charged slepton sector is a deeper probe than the chargino/neutralino or
Higgs sectors of the flipped $SU(5)$ model at LEPII, while the opposite is true
for the minimal $SU(5)$ model where the slepton sector is no probe at all.
\Date{}

\newsec{Introduction}
The quest for a theoretical understanding of supersymmetry and its
phenomenological consequences has being going on for over a decade. So far
no supersymmetric particle has been directly observed in accelerator
experiments or indirectly in proton decay or dark matter detectors. However,
the recent precise LEP measurements of the gauge coupling constants can be
taken in the context of supersymmetric grand unification as indirect evidence
for virtual supersymmetric corrections \EKN. This observational situation may
appear discouraging to some. However, it really should not since from a totally
unbiased point of view, most sparticle masses could lie anywhere up to a few
TeV, with no particular correlations among them. This means that existing
facilities (Fermilab, LEPI,II, HERA, Gran Sasso) as well as future ones
(LHC, SSC) are needed in order to truly explore the bulk of the supersymmetric
parameter space.

On the other hand, specific supergravity models incorporating well motivated
theoretical constraints can be very predictive, and perhaps even fully tested
in the next few years with the present generation of collider experiments at
Fermilab, HERA, and LEPII. We have recently focused
our attention on two such models: (i) the minimal $SU(5)$ supergravity model
including the severe constraints of proton decay
\refs{\MATS,\ANabc,\LNP,\HMY,\LNPZ} and a not too young Universe
\refs{\LNZa,\LNP,\LNPZ,\ANdm},
and (ii) a recently proposed no-scale flipped $SU(5)$ supergravity model \LNZb.
The parameter spaces of these models have been scanned and a set of allowed
points has been identified in each case. Several results then follow for the
sparticle masses. These are summarized in Table I and discussed in detail
in Refs. \refs{\LNZa,\LNP,\LNPZ,\LNWZ,\LNPWZh} for the minimal $SU(5)$ model
and in Refs. \refs{\LNZb,\LNWZ,\LNPWZh} for the flipped model. As far as the
sparticle masses are concerned, perhaps the most striking difference between
the two models is in the slepton masses which are below $\approx300\GeV$ in
the flipped $SU(5)$ model, while they are out of reach of existing facilities,
\ie, above $300\GeV$, in the minimal $SU(5)$ model. The study of the
specific models like the two we are pursuing, singles out small regions of the
vast 21-dimensional parameter space of the MSSM (minimal supersymmetric
extension of the Standard Model). We have already shown \refs{\LNPZ,\LNWZ} that
experimental predictions for these models can be so precise that potential
discovery or exclusion in the next few years is a definite challenge.

In a previous paper \LNWZ\ we have studied the prospects for supersymmetry
detection at Fermilab in the neutralino-chargino sector. Here we continue
our general program by exploring the supersymmetric signals for charginos,
neutralinos, sleptons, and the lightest Higgs boson at LEPII in the two models.
For charginos we study the reaction $e^+e^-\to\chi^+_1\chi^-_1$ and the
subsequent `mixed' (1 lepton plus 2 jets plus $\slash{p}$) and dilepton decay
signatures. We show that the predicted number of mixed events for both models
are experimentally significant up to the kinematical limit, and do not
overlap. Therefore, if $m_{\chi^\pm_1}<100\GeV$, then LEPII should be
able to exclude at least one of the models.
For neutralinos we analyze $e^+e^-\to \chi^0_1\chi^0_2$ and the dilepton
signature, as a means to indirectly probe chargino masses above $100\GeV$.
The charged slepton sector appears very interesting for LEPII in the
predictions of the flipped $SU(5)$ supergravity model. We compute the number of
dilepton events expected from pair-produced
$\tilde e\tilde e,\tilde\mu\tilde\mu,\tilde\tau\tilde\tau$, and conclude
that these also should be accessible up to the kinematical limit.
Finally, we explore the Higgs sector and study $e^+e^-\to Z^*h$ production, the
branching ratios $h\to b\bar b,\tau^+\tau^-,c\bar c,gg$ and the `invisible'
mode $h\to\chi^0_1\chi^0_1$. We show that the latter can have a branching ratio
as large as $30\%$, therefore significantly eroding the preferred
$h\to2\,{\rm jets}$ mode. Nonetheless, detection is possible in a large
fraction of parameter space for both models at LEPII. Throughout this paper we
emphasize the interconnections among the various sectors of the models and
their experimental consequences. For example, charged slepton pair production
should indirectly probe chargino masses as high as $150\GeV$ in the flipped
model.

\newsec{Charginos and Neutralinos}
Among the various supersymmetric neutralino/chargino production processes
accessible at LEPII, the one with the largest cross section is
$e^+e^-\to\chi^+_1\chi^-_1$ which proceeds through $s$-channel $\gamma^*$- and
$Z^*$-exchange and $t$-channel $\tilde\nu_L$-exchange. This cross section has
been calculated in the literature for various limiting cases of the chargino
composition and for a general composition (\ie, an arbitrary linear combination
of wino and charged higgsino components) as well. We have independently
calculated the
cross section in the general case, and our result agrees with \eg, Ref. \BFM.
The cross sections for this process for both models are shown in Fig. 1 for
$\sqrt{s}=200\GeV$. The reason the cross sections are lower in the no-scale
flipped model is due to a well known destructive interference between the $s$-
and $t$-channels, which is relevant for light $\tilde\nu_L$ masses, or more
properly for $m_{\tilde\nu_L}\sim m_{\chi^\pm_1}$. In addition, for $|\mu|\gg
M_2\approx0.3m_{\tilde g}$ ($M_2$ is the $SU(2)_L$ gaugino mass)
the $\chi^\pm_1$ mass eigenstate is predominantly gaugino and therefore
its coupling to lepton-slepton is not suppressed by the small lepton masses.
In the minimal $SU(5)$ model, $m_{\tilde\nu_L}>500\GeV$ and the contribution of
the $t$-channel is small. In the flipped model $m_{\tilde\nu_L}\sim
m_{\chi^\pm_1}$ and the destructive interference is manifest.

The best signature for this process is presumed to be the one-charged lepton
($e^\pm$ or $\mu^\pm$) + 2-jets + $\slash{p}$ or `mixed' mode, where one
chargino
decays leptonically and the other one hadronically \refs{\Dionisi,\Tata}. In
the minimal $SU(5)$, since the sleptons and squarks are heavy, the chargino
decays are mediated dominantly by the $W$-exchange channels \LNWZ\ and one gets
${\rm BR}(\chi^\pm_1\to\chi^0_1 l^\pm\nu_l)_{minimal}\approx2/9$ ($l=e+\mu$)
and ${\rm BR}(\chi^\pm_1\to\chi^0_1 q \bar q')_{minimal}\approx2/3$.

For the flipped case things are more complicated due to the light
slepton-exchange channels. There are three regimes which one can identify: (i)
when the slepton exchange channels dominate, the leptonic branching ratio
(into $l=e+\mu$) is
$\approx2/3$ and the hadronic one goes to zero; (ii) when the $W$-exchange
channels dominate (as in the minimal $SU(5)$ case), the leptonic branching
ratio drops down to $\approx2/9$ and the hadronic one grows up to $\approx2/3$;
and (iii) in the transition region between these two regimes, destructive
interference between the $W$-exchange and slepton-exchange amplitudes can
suppress the leptonic branching ratio and enhance the hadronic one beyond
their values at the end of the transition. In Fig. 2 we show an example of
this phenomenon for $m_t=100\GeV$; for larger values of $m_t$ the effect is
less pronounced (see Fig. 2 in Ref. \LNWZ).

In Fig. 3 we show the number of `mixed' events to be expected per ${\cal
L}=100\ipb$ for both models, \ie,  $\sigma(e^+e^-\to\chi^+_1\chi^-_1)\times
{\rm BR}(\chi^+_1\to \chi^0_1l^+\nu_l)\times{\rm BR}(\chi^\pm_1\to
\chi^0_1q\bar q')\times 2$, where the factor of two accounts for summing over
the two
charges of the outgoing lepton. The very small numbers for the flipped model
which occur mostly for $\mu>0$ correspond to points in the parameter space
where the slepton-exchange channels dominate the chargino decays and the
hadronic branching ratio goes to zero (case (i) in the previous paragraph).
Perhaps the most interesting feature of these results is that the
predicted number of events for both models {\it do not overlap}. Therefore,
{\it if} $m_{\chi^\pm_1}<100\GeV$, then LEPII should be able to {\it exclude}
at least one of the models (and possibly even both).

As far as the backgrounds are concerned, the dominant one is $e^+e^-\to W^+W^-$
with one $W$ decaying leptonically and the other one hadronically. (To a lesser
extent, the $f\bar f(\gamma),Ze(e),W\nu(e),ZZ$, and $Z\nu\nu$ backgrounds
also apply \Felcini.) Several features of the chargino decays (such as an
isolated lepton, missing mass, hadronic mass, etc.) allow for suitable cuts to
be made which reduce the $WW$ background to very small levels
\refs{\Tata,\Felcini,\GrivazI}. Model-dependent studies indicate that a
$5\sigma$ effect (\ie, $S/\sqrt{B}\ge5\sigma$) can be observed with
${\cal L}=100\,(500)\ipb$ of integrated luminosity for
$\sigma(e^+e^-\to\chi^+_1\chi^-_1)\gsim0.40\,(0.17)\pb$
\GrivazI. These calculations assume $W$-exchange dominance in chargino decays
(case (ii) above) and are therefore applicable to the minimal $SU(5)$ model.
In this case Fig. 3 shows that one could explore all allowed points in
parameter space with $m_{\chi^\pm_1}<100\GeV$, since
$\sigma\gsim0.40\,(0.17)\pb$ for ${\cal L}=100\,(500)\ipb$ would require
$40\,(85)$ observed events. Moreover, since in this model
$m_{\chi^\pm_1}<104\,(92)\GeV$ for $\mu>0\,(\mu<0)$ \LNPWZh, only a few points
in parameter space should remain unexplored in this direct way at LEPII.

For the flipped model the experimental study referred to above may not apply
since the $W$-exchange dominance assumption is not likely to hold for
$m_{\chi^\pm_1}<100\GeV$ (see \eg, Fig. 2). Assuming that the results apply
at least approximately, we can see that a good fraction of the parameter
space for $m_{\chi^\pm_1}<100\GeV$ could be explored.
If no signal is observed, this would imply that $m_{\chi^\pm_1}>100\GeV$
in the minimal $SU(5)$ model, but not necessarily in the flipped model because
of possible highly suppressed hadronic decay channels.
To probe the remaining unexplored regions of the flipped model for
$m_{\chi^\pm_1}<100\GeV$ we show in Fig. 4 the predicted number of events
for $e^+e^-\to\chi^+_1\chi^-_1\to{\rm dileptons}$ which does not suffer
from small chargino hadronic branching ratios. However, the efficiency cut
needed to suppress the backgrounds to this process is not known at present.
Nevertheless, the signal is quite significant and should encourage experimental
scrutiny.\foot{In the next section we show that the chargino-dilepton-signal
is in general a `background' to dileptons from charged slepton decays.
Therefore experimental isolation of the chargino-dilepton-signal may be
required anyway.} For the minimal $SU(5)$ case the dilepton signal is $1/6$ of
that for the mixed mode: a factor of 3 is lost in substituting the hadronic
branching fraction ($2/3$) for the leptonic one ($2/9$), and a further factor
of 2 from not needing to sum over the charges of the outgoing lepton. Thus,
Fig. 3 (top row$\times{1\over6}$) shows that the minimal model dilepton signal
is much smaller than the flipped model one.

Concerning neutralino detection at LEPII, the largest observable cross section
occurs for $e^+e^-\to\chi^0_1\chi^0_2$ which is mediated by $Z^*$ $s$-channel
exchange and $\tilde e_{L,R}$ $t$-channel exchange. Since in the models we
consider $m_{\chi^\pm_1}\approx m_{\chi^0_2}\approx2m_{\chi^0_1}$, this
process could explore indirectly chargino masses up to $\sim130\GeV$ and may
be worth considering despite the potentially small rates. The coupling
$Z\chi^0_1\chi^0_2$ depends exclusively on the higgsino admixture of $\chi^0_1$
and $\chi^0_2$ and is thus highly suppressed here (and so is the $s$-channel
amplitude) and in any model where $|\mu|\gg M_2$. The cross section then
depends crucially on the $t$-channel amplitude, \ie, on the selectron mass.
For the minimal $SU(5)$ model we find $\sigma(e^+e^-\to\chi^0_1\chi^0_2)<10\fb$
since $m_{\tilde e_{L,R}}>500\GeV$. Even with $100\%$ efficiencies and high
branching ratios, it would take at least ${\cal L}=1000\ipb$ to get an
observable
signal at the largest cross section. For $m_{\chi^\pm_1}>100\GeV$ the cross
section drops below $0.1\fb$ and therefore this mode is hopeless for
exploration of chargino masses above $100\GeV$ at LEPII in the minimal $SU(5)$
model.

For the flipped model we have $m_{\tilde e_R}<190\GeV$ and $m_{\tilde
e_L}<300\GeV$ and thus the cross section for $e^+e^-\to\chi^0_1\chi^0_2$
is correspondingly much larger, although slightly below $1\pb$ at most; see
Fig. 5 top row. With ${\cal L}=500\ipb$ and a neutralino dilepton branching
ratio as high as $2/3$ (see Fig. 4 in Ref. \LNWZ), one could get an observable
number of neutralino-dilepton ($\chi^0_2\to\chi^0_1l^+l^-$) events even for
$m_{\chi^\pm_1}>100\GeV$; see Fig. 5
bottom row.\foot{Note that neutralino-dileptons need still to be distinguished
from the chargino-dileptons discussed above; the one-sided nature of the former
signal may help in this regard.} Most of the backgrounds to
this process can be reduced, except for the $WW$ one which was shown in
Ref. \Tata\ to overwhelm the signal for both leptonic and hadronic decays,
at least for the parameters considered by those authors.
A re-evaluation of this analysis in the light of the flipped model cross
section and branching ratios would need to be performed to be certain of the
fate of this mode. Since this is one way in which LEPII could indirectly
explore chargino masses above $100\GeV$, it would appear to be a worthy
exercise.

\newsec{Sleptons}
The charged sleptons ($\tilde e_{L,R},\tilde\mu_{L,R},\tilde\tau_{L,R}$)
offer an interesting supersymmetric signal through the dilepton decay mode,
if light enough to be produced at LEPII \refs{\Dionisi,\Tata}. This is
partially the case for the flipped $SU(5)$ model where $m_{\tilde e_L}\lsim
300\GeV$ and $m_{\tilde e_R}\lsim200\GeV$. (No such signal exists at LEPII for
the minimal $SU(5)$ model since $m_{\tilde l}>300\GeV$.) We have computed the
cross sections for
\eqna\SI
$$\eqalignno{e^+e^-\to&\tilde e^+_L\tilde e^-_L,\tilde e^+_R\tilde e^-_R,
\tilde e^\pm_L\tilde e^\mp_R,&\SI a\cr
e^+e^-\to&\tilde \mu^+_L\tilde \mu^-_L,\tilde \mu^+_R\tilde \mu^-_R,&\SI b\cr
e^+e^-\to&\tilde \tau^+_L\tilde \tau^-_L,\tilde \tau^+_R\tilde \tau^-_R.&\SI
c\cr}$$
The $\tilde e^+_L\tilde e^-_L,\tilde e^+_R\tilde e^-_R$ final states receive
contributions from $s$-channel $\gamma^*$ and $Z^*$ exchanges and $t$-channel
$\chi^0_i$ exchanges, while the $\tilde e^\pm_L\tilde e^\mp_R$ only proceeds
through the $t$-channel. The $\tilde \mu^+_L\tilde \mu^-_L,\tilde \mu^+_R\tilde
\mu^-_R$ and $\tilde \tau^+_L\tilde \tau^-_L,\tilde \tau^+_R\tilde \tau^-_R$
final states receive only $s$-channel contributions, since all couplings are
lepton flavor conserving, and therefore mixed LR final states are not allowed
for smuon or stau production. Our results agree with those in Ref. \BFMsf. Note
that in the flipped model the left-handed (L) slepton masses are considerably
heavier than the right-handed (R) ones (see Fig. 3 in Ref. \LNZb). In
particular $m_{\tilde\tau_R}<m_{\tilde e_R,\tilde\mu_R}<m_{\tilde
e_L,\mu_L}<m_{\tilde\tau_L}$.

The acoplanar dilepton signal associated with selectron pair production
has been traditionally assumed to come entirely from $\tilde e^\pm_{L,R}\to
e^\pm\chi^0_1$ decay channels, \ie, purely di-electrons. This is an
idealization which need not hold in specific supergravity models. In the
flipped model the following decay channels are allowed:
\eqna\SII
$$\eqalignno{\tilde e^\pm_L\to&
e^\pm\chi^0_1,e^\pm\chi^0_2,\nu_e\chi^\pm_1,&\SII a\cr
\tilde e^\pm_R\to& e^\pm\chi^0_1,e^\pm\chi^0_2,&\SII b\cr}$$
If $\chi^0_2$ decays invisibly ($\chi^0_2\to\nu\bar\nu\chi^0_1$) and
$\chi^\pm_1$ leptonically ($\chi^\pm_1\to l^\pm\nu_l\chi^0_1$, $l=e,\mu,\tau$),
then one has new contributions to the sought-for dilepton signal. Note however
that in the latter case the charged leptons ($e,\mu,\tau$) will be
`non-leading' and thus their spectrum (from three-body $\chi^\pm_1$ decay)
is likely to deviate from the `leading' lepton spectrum (from two-body $\tilde
e_{L,R}$ decay). Because of the details of the model, more than
90\% of the points in the allowed parameter space have $m_{\tilde
e_R}<m_{\chi^0_2}$ and therefore ${\rm BR}(\tilde e^\pm_R\to
e^\pm\chi^0_1)=100\%$ for these points; the remaining points, which allow
$m_{\tilde e_R}>m_{\chi^0_2}$, have branching fractions to $e^\pm\chi^0_1$ no
smaller than 75\%. On the other hand, for all points in parameter space we
find $m_{\tilde e_L}>m_{\chi^0_2,\chi^\pm_1}$ and the decays of the heavier
$\tilde e_L$ proceed in all three ways.

It is important to realize that in this model, for a given point in parameter
space, all slepton masses are determined, and the lighter of the final states
in Eq. \SI{a} ($\tilde e^+_R\tilde e^-_R$) will dominate the total cross
section into selectron pairs. Moreover, the dilepton signal from this
dominant contribution will be purely leading di-electrons. The other final
states in Eq. \SI{a} involving the heavier $\tilde e_L$ ($\tilde e^\pm_L\tilde
e^\mp_R,\tilde e^+_L\tilde e^-_L$) have smaller cross sections and contribute
mostly leading di-electrons. This is because non-leading leptons ($e,\mu,\tau$)
require the production of the heavier $\tilde e^\pm_L$ and the further
branching ratio suppressed decay into $\nu_l\chi^\pm_1$. Therefore,
we expect the traditional  acoplanar di-electron $+\slash{p}$ (missing energy)
signature to prevail.

We have computed the total (leading) di-electron signal (per ${\cal
L}=100\ipb$) from
all channels in Eq. \SI{a}. The result is shown in Fig. 6 (top row) as a
function of the selectron mass $m_{\tilde e_R}$. The thinning of the
point distributions for $m_{\tilde e_R}\gsim80\GeV$ is due to the kinematical
closing of the $\tilde e^\pm_L\tilde e^\mp_R$ production channel. The number of
di-electron events is quite significant and with adequate experimental
efficiencies to account for the di-electron $WW,ZZ$ decay backgrounds
\refs{\Felcini,\GrivazI,\Kat} it should be possible to explore the whole
kinematically allowed mass range (\ie, $m_{\tilde e_R}\lsim100\GeV$ and
indirectly larger $m_{\tilde e_L}$ masses) with ${\cal L}=500\ipb$. For
example, a study in Ref. \Felcini\ indicates that one would
need $\sigma(e^+e^-\to\tilde e\tilde e)\gsim0.1\pb$ to observe a $5\sigma$
effect. From Fig. 6 this will allow exploration up to $m_{\tilde e_R}\approx
95\GeV$.

The analysis in the previous paragraphs for selectron pair production can be
carried over to $\tilde\mu$ and $\tilde\tau$ pair production. In this case
$\tilde\mu^+_R\tilde\mu^-_R$ and $\tilde\tau^+_R\tilde\tau^-_R$ dominate the
production cross section and the leading di-muons and di-taus  constitute
the bulk of the dilepton signal respectively. Furthermore, non-leading
leptons ($e,\mu,\tau$) are even less likely to occur here since the larger
contributions from $LR$ final states (compared to $LL$ final states) in the
selectron case, are not present here. In Fig. 6 (bottom row) we show the result
for the di-tau case; the di-muon signal is very similar. These signals are
smaller (although not by too much) than the di-electron signal in selectron
pair production because of the fewer production diagrams.

The slepton dilepton signal could also be used to explore indirectly values
of the chargino masses beyond the direct kinematical limit of 100 GeV. In
Fig. 7 (top row) we show the $\tilde e$ di-electron signal versus the chargino
mass, and observe that in principle one could probe as high as
$m_{\chi^\pm_1}\approx150\GeV$. In fact, this indirect method appears to be
much more promising than the one suggested in Sec. 2 through the
$\chi^0_1\chi^0_2$ channel.

It is important to realize that dileptons also occur in chargino (and to
a lesser extent $\chi^0_2$) decays, as
discussed for the flipped model in Sec. 2. In Fig. 7 (bottom row) we show the
number of chargino-dileptons from Fig. 4 but this time plotted against
$m_{\tilde e_R}$. This `background' to slepton-dileptons (the search topology
for dileptons is the same) has some features which may allow for it to be
sufficiently accounted for. The slepton-dileptons contain only (leading)
$l^+l^-$ ($l=e,\mu$ for now) pairs, whereas the chargino-dileptons in Fig. 7
contain a mixture of $25\%$ $e^+e^-$, $25\%$ $\mu^+\mu^-$, and $50\%$
$e^\pm\mu^\mp$. Moreover, the common $l^+l^-$ pairs have a different energy
spectrum ({\it c.f.} $\tilde l^\pm\to l^\pm\chi^0_1$ with $\chi^\pm_1\to
l^\pm\nu_l\chi^0_1$). Of course, if charginos are observed through the mixed
signal, one could simply `subtract out' the ensuing chargino dilepton signal
from the total observed (chargino+slepton) dilepton signal.

The previous two paragraphs exemplify the interconnections among the various
sectors of this class of models. These correlations allow experimental
exploration of one sector to probe indirectly other sectors. They also allow
for a reliable computation of all contributing sectors to a particular physics
signal (such as acoplanar di-electrons).

We have not considered the production of sneutrinos since because of their
masses (in between the R and L charged lepton masses) the rates will be lower
than for selectron production. Moreover, their visible decay channels
($\tilde\nu_e\to\nu_e\chi^0_2,e^\pm\chi^\pm_1$) are branching-ratio suppressed
(since $\tilde\nu_e\to\nu_e\chi^0_1$ is expected to dominate), and will lead to
one-sided (likely soft) dileptons.

\newsec{The lightest Higgs boson}

We now consider the standard alternative to direct sparticle production and
decay modes, namely the SUSY Higgs sector. Since the supergravity models we
consider here contain two complex Higgs doublets, after spontaneous
symmetry breaking the physical Higgs spectrum contains the usual $h,H$
($CP$-even), $A$ ($CP$-odd) neutral Higgs fields, and the charged
$H^\pm$ Higgs field. For a comprehensive review of the SUSY Higgs sector,
we refer the reader to Ref. \HHG. Our goal in this section is to reformulate
the `generic' analysis for SUSY Higgs production and decay in terms of the
{\it specific} minimal  and flipped $SU(5)$ supergravity models described
above. As a result we must include some non-standard decay channels, such as
$h\rightarrow\chi^0_1 \chi^0_1$, which are usually not considered in generic
analyses since they are so model dependent. Significantly,
${\rm BR}(h\rightarrow\chi^0_1 \chi^0_1)$ can be quite large in these models,
and this modifies the usual assumptions regarding Higgs signals at colliders.
We also incorporate the one-loop corrections to the Higgs masses which can be
quite significant in large regions of parameter space \ERZ.

{}From the underlying radiative breaking mechanism in the two supergravity
models we consider, and the experimental lower bound on the gluino mass, one
can show \LNPWZh\ that the Higgs sector of both models approaches
a SM-like situation with a light scalar ($h$) with SM-like couplings, and a
heavy Higgs spectrum ($H,A,H^\pm$) which tends to decouple from fermions and
gauge bosons for increasing
$m_{\tilde g}$ (see Ref. \LNPWZh\ for further details). The approach to this
limit is accelerated (as a function of $m_{\tilde g}$) in the minimal $SU(5)$
model due to the proton decay constraint which requires large scalar masses.
In the flipped model the $A$-Higgs can  be relatively light for large
$\tan\beta$ ($m_A\gsim m_h$), implying a slower approach to the limiting
situation. For the most part then, we need only consider the $h$ Higgs at LEP.
Thus, the phenomenological analysis of the Higgs sector for each supergravity
model simplifies dramatically, particularly when making contact with previous
experimental and Monte-Carlo results. With proper care for the non-standard
$hZZ$ coupling and branching ratios (which simply amounts to a re-scaling of
the SM analysis), we can adopt the definitive SM analysis of Ref. \Wu\ for
$m_{H_{SM}}\lsim 80 \GeV$ along with the recent results summarized in Ref.
\Kat\ for $m_{H_{SM}}\gsim 80 \GeV$.

With regard to Higgs production, since we take $\sqrt{s}=200\GeV$,
the only relevant mode is the standard $s$-channel
$e^+e^-\rightarrow Z^* h\rightarrow h f\bar f$ production process, since the
$Hf\bar f,H^+H^-$ final states are kinematically forbidden, and the $WW$-fusion
$t$-channel processes are relevant only for $\sqrt{s}\gsim 400\GeV+0.6m_h$
\HHG. For the flipped case, there are a few exceptional ($<1\%$) points in the
allowed parameter space for which the associated $e^+e^-\rightarrow hA$ process
is kinematically allowed (\ie, for $m_h\lsim m_A\lsim90\GeV$ which is possible
for large $\tan\beta$ values, see Fig. 6 in Ref. \LNZb); we neglect this mode
in our analysis. In Fig. 8
we show the cross section $\sigma(e^+e^-\rightarrow Z^* h\rightarrow h \nu \bar
\nu)$ vs. $m_h$ for both models for $\sqrt{s}=200\GeV$. The values shown for
the minimal $SU(5)$ model also correspond to the SM result since one can
verify that $\sigma(e^+e^-\to Z^*h)/\sigma(e^+e^-\to Z^*
H_{SM})=\sin^2(\alpha-\beta)>0.9999$ in this case \LNPWZh. As a reference
point, for $m_h=80\GeV$, and the canonical integrated luminosity
of $500 {\rm pb}^{-1}$, the SM (and also the minimal $SU(5)$) value of
$\sigma\simeq 0.145\pb$ would correspond to approximately $62$ $b\bar b\nu\bar
\nu$ events ($56$ if initial state radiation is included) \Wu. The `tail' in
the cross section when the second $Z$ is forced off-shell is evident in the
flipped case for $m_h\gsim 110 \GeV$. In this model, significant deviations
from the SM curve are possible (see Fig. 8 and compare top and bottom rows),
and these correspond to the small set of points ($<1\%$) for which $m_A$ can
achieve moderately light values ($m_A\gsim m_h$ for large $\tan\beta$). Since
in this case $\sin(\alpha-\beta)\gsim 0.8$, a reduction of up to $\approx1/3$
in the cross section is possible compared to the SM case. We should add that
these results agree quite well with previous results obtained in Ref. \KLNPYh\
for a slightly different version of the no-scale flipped $SU(5)$ supergravity
model. There the reduction in $\sin^2(\alpha-\beta)\sim0.8$ is only possible
for large $\tan\beta$ values, and $80\GeV\lsim m_h\lsim100\GeV$. In addition,
in Ref. \KLNPYh\ it was found that $m_h\lsim120\GeV$, and this limit is close
to the one obtained in the present version of the flipped model
($m_h\lsim135\GeV$).

Regarding the decay of the $h$ Higgs, it is well-known that higher order
QCD corrections can be important and affect the overall results
for the relevant decay channels $h\rightarrow b\bar b,c\bar c$
by up to $20\%$ \QCD. This should be particularly true at LEPII, where
$m_h$ is much greater than $m_{c,b}$ and there will be significant
running of the quark masses from the scale $Q=m_{c,b}$ up to $Q=m_h$.
We therefore choose to include QCD corrections to
${\cal O}(\alpha^2_s)$ as outlined in detail in Ref. \Barger.
In addition, significant departures from the usual SM
branching ratios result when we include the $h\rightarrow \chi^0_1\chi^0_1$
decay channel. Statistically speaking, we find that $7\%(12\%)$ of the points
for the flipped model for $\mu>0$ ($\mu<0$) kinematically allow for this
non-standard neutralino decay mode. In the minimal $SU(5)$ case, this
fraction rises to $26\%(36\%)$ for $\mu>0$ ($\mu<0$). \foot{There are only a
handful of points for which the $h\rightarrow \chi^0_1\chi^0_2$ mode is
kinematically accessible
in the flipped model (there are none in the minimal $SU(5)$ model) with
${\rm BR}(h\to\chi^0_1\chi^0_2)\lsim10^{-4}$. We do not show these points
here.} In Fig. 9 and Fig. 10 we show the branching ratios for $h\rightarrow
b\bar b$, and $\chi^0_1 \chi^0_1$ respectively for both models. For
${\rm BR}(h\rightarrow b\bar b)$, the `standard' expectation of $\approx 0.85$
is evident, however the value can be reduced down to $0.6\;(0.5)$ in the
minimal(flipped) $SU(5)$ model predominantly due to the opening of the
$\chi^0_1\chi^0_1$ channel. (The few
points for $\mu<0$ in the minimal model which have yet smaller branching ratios
correspond to an enhancement of the $h\to gg$ rate, as described below.) In the
flipped case, the $h\to\chi^0_1\chi^0_1$ mode is maximized for
$m_h\approx105\GeV$ (see Fig. 10 bottom row), while for the minimal case
this happens for $m_h\gsim 75\GeV$ (see Fig. 10 top row).

As for the other decay channels ($h\rightarrow gg,\tau^+\tau^-,c\bar c$),
in Fig. 11 we show the ${\rm BR}(h\rightarrow gg)$ which varies considerably
over the parameter space, and generally increases with $m_t$.\foot{We use the
expressions for $h\rightarrow gg$ that appear in Ref. \Weiler\ along with minor
corrections pointed out by those authors.} The three packs of curves in Fig. 11
for the flipped model (bottom row) correspond from left-to-right to
$m_t=100,130,160\GeV$; the further structure is due to the $\tan\beta$ values
used which increment in steps of two units starting at $\tan\beta=2$ and
accumulate for large $\tan\beta$. For the flipped model we find ${\rm BR}(h\to
gg)\lsim 0.2$. This is generally true for the minimal model also, except for
some extreme cases for large $m_t\simeq 160\GeV$, when it can be as large as
$\approx 0.9$. This enhancement is due to a large $\tilde t_{L,R}$ mixing
(which occurs in a small region of parameter space), which results in a {\it
very} light stop mass $m_{\tilde t_1}\approx
50\GeV$ and therefore suppresses the $b\bar b$ mode significantly (note the
points with uncharacteristically small values of ${\rm BR}(h\to b\bar b)$ in
Fig. 9 top row for $\mu<0$). We expect ${\rm BR}(h\rightarrow \gamma\gamma)$
to be much smaller, and do not include this mode in our analysis here.

Although not shown, we have calculated ${\rm BR}(h\rightarrow \tau^+\tau^-)$
also, and find that for both models, this channel is confined to a narrow
band centered around  ${\rm BR}(h\rightarrow \tau^+\tau^-)=0.08$; this agrees
well with the results obtained in Refs. \refs{\Barger,\Kunszt}.
In Fig. 12 we show ${\rm BR}(h\rightarrow c\bar c)$ for both models. For the
flipped case, there is a large range of values $(0.0001\lsim{\rm
BR}(h\rightarrow c\bar c) \lsim 0.06)$, with a noticeable dip in the range
$80\GeV\lsim m_h\lsim 100\GeV$ corresponding to the deviation from the SM
couplings. For the minimal $SU(5)$ case,
${\rm BR}(h\rightarrow c\bar c)\sim 0.06$. The latter result is predominantly
due to the fact that in the minimal case $\sin(\alpha-\beta)\approx1$ and
therefore the $h$-$c$-$\bar c$ coupling ($\propto\cos\alpha/\sin\beta$) goes to
the SM $H_{SM}$-$c$-$\bar c$ coupling, for virtually all points.

Detectability of the $h$ Higgs requires the combination of production and
experimentally important decay modes, as well as a detailed treatment of
the backgrounds and overall efficiency.\foot{In what follows we adapt the
results of Ref. \Wu\ for the SM Higgs to the $h$ Higgs for $m_h\lsim 80 \GeV$.}
{}From our previous discussion of $h$ production and decay, it is clear that
the fraction of $h$ Higgs events compared to the SM will be
\eqn\RR{R_h\equiv \sin^2(\alpha-\beta)\cdot f,}
where $f\equiv {\rm BR}(h\to X)/{\rm BR}(H_{SM}\to X)$, and $X$ is a specific
Higgs final state. As for the backgrounds,
the various SM $e^+e^-\rightarrow ZZ,W^+W^-,Ze^+e^-,We\nu_e,q\bar q\gamma$
modes apply to a different degree depending on the particular production
channel. For the $(h\to jj)\nu \bar \nu$ final states we consider here
($j$=jet), the $ZZ,W^+W^-,q\bar q\gamma,We\nu_e$ backgrounds are dominant.
Considering the SM analysis first, Ref. \Wu\ finds that for
$m_{H_{SM}}\simeq 80\GeV$, the efficiency ($\epsilon$) is $\sim 21\%$ for the
dominant $e^+e^-\to H_{SM} Z^*\to (H_{SM}\to b\bar b)\nu \bar \nu$ channel.
This corresponds to
\eqna\cross
$$\eqalignno{\sigma(e^+e^-\to Z^*H_{SM}\to\nu\bar\nu H_{SM})\cdot&\,{\cal L}
\cdot{\rm BR}(H_{SM}\to b\bar b)\cdot\epsilon\cdot\epsilon_{ISR}=\cr
&=(0.145)\cdot(500)\cdot(0.85)\cdot(0.21)\cdot(0.91)\approx12&\cross {}}$$
expected events ($\epsilon_{ISR}$ accounts for the initial state radiation
effects at $m_{H_{SM}}=80\GeV$) with a background of 4 events, leading to a
signal/background of $\simeq 3$. For $m_{H_{SM}}\gsim 80\GeV$, both the signal
and efficiency decrease further. Thus, $m_{H_{SM}}\simeq 80\GeV$ has until
recently been considered the limit of detectability for the SM Higgs.

Turning now to the supersymmetric Higgs analysis, for $m_h<80 \GeV$, we find
that $R_h\gsim 0.7$ for both models, where $f={\rm BR}(h\to b\bar b+c\bar
c+gg)/{\rm BR}_{SM}$. Thus, additional integrated luminosity would be needed in
order to probe up to $m_h\simeq 80 \GeV$ from two-jet reconstruction off the
$Z$ via $hZ^*$ production. Despite the degradation of the favored two-jet
signal, for $m_h\lsim 80 \GeV$ detection via recoil against $l^+l^-$ pairs or
2-jets may still be possible through the
$e^+e^-\to h Z^*\to(h\to X)l^+l^-,(h\to X)jj$ channels, where $X$ is invisible.
Overall, a detailed Monte-Carlo study would be needed to determine the
experimental mass limits for a Higgs decaying invisibly.

The analysis summarized in Ref.
\Treille\ has demonstrated that with $b$-quark tagging from $H_{SM}\to b\bar b$
(and for $\sqrt{s}=190 \GeV$), a $90 \GeV$ SM Higgs should be detectable above
background with ${\cal L}=300\ipb$ luminosity at the $5 \sigma$ level \Kat.
This extends
the experimental Higgs reach at LEP even further, through the previously
troublesome region where $m_{H_{SM}}\sim M_Z$. For higher beam energies, the
Higgs mass reach is expected to be $\sqrt{s}-100 \GeV$. If the beam energy
could be pushed up to the magnet limit of $\sqrt{s}=240 \GeV$, a value of
$m_{H_{SM}}\simeq 140 \GeV$ could in principle be explored, however a reach of
$m_{H_{SM}}\lsim 100 \GeV$ is more realistic for the near future, corresponding
to $\sqrt{s}=200 \GeV$.

For $m_h\gsim 80 \GeV$, we define $f\equiv{\rm BR}(h\to b\bar b)/{\rm BR}_{SM}$
since $b-$tagging is the only source of the signal. For both models we find the
value of $R_h$ can be as small as $\sim 0.60$ for the experimentally accessible
region $m_h\lsim 100 \GeV$.\foot{We exclude from the discussion the very few
points for $\mu<0$ in the minimal $SU(5)$ model where ${\rm BR}(h\to
gg)\approx0.9$ and the $f$-ratio drops to values as low as $0.25$.} In this
case, the irreducible background is certain to obscure the $50-80\%$ reduction
in signal in a $b-$tagging analysis.

Thus, we conclude that for the minimal $SU(5)$ model and for $\mu>0$, the $h$
Higgs will most likely be seen at LEP200 since $m_h\lsim 83 \GeV$ and
$R_h\gsim0.7$. For $\mu<0$ in the minimal model and for both signs of $\mu$
in the flipped model, the $h$ Higgs could escape detection at LEP200 if $80\GeV
\lsim m_h \lsim 100\GeV$ and $R_h$ is not big enough, \ie, when the
non-standard $hZZ$ coupling along with the reduction of ${\rm BR}(h\rightarrow
b\bar b)$ due to $h\rightarrow \chi^0_1\chi^0_1$ is significant, and/or if the
$\tilde t_1$ is very light, and $h\rightarrow gg$ would overwhelm all other
channels. It is of course possible that for this small set of points ($<1\%$),
the associated Higgs production processes $e^+e^-\to hA\to b\bar b b\bar
b,b\bar b\tau^+\tau^-$ may open up and allow for Higgs detection \Treille. We
must conclude that in the flipped model that we have considered here, it is
possible (but unlikely) that nature could conspire to fall within the so-called
`tie' region in the $(\tan\beta,m_A)$ plane where neither process could be seen
at LEPII. In this unlikely event, the $h$ (and/or the $A$ Higgs) could
conceivably escape detection. (For the minimal model $m_A\gg M_Z$ and the `tie'
region is avoided entirely.) For the flipped model (and for light $m_A\gsim
m_h$) in the mass region $m_h>80\GeV$, the only possible hope would be looking
for the $h$ at a $500\GeV\;e^+e^-$ machine or at the SSC/LHC
\refs{\GrivazI,\BargerII}.

The present lower bound for the $h$ Higgs is $m_h>43\GeV$ \GrivazII. This limit
is regarded as {\it model-independent}, valid for $m_{\tilde q}<1\TeV$, and
assumes SM final state products. In the models we consider here, we have shown
that the $h\to\chi^0_1\chi^0_1$ mode should also be considered for some regions
of parameter space. One can see however from Fig. 9 that for $m_h\lsim 43\GeV$
the $h\to\chi^0_1\chi^0_1$ mode is relatively unimportant. Even for
$m_h\lsim 60\GeV$, the non-standard reduction of
${\rm BR}(h\to b\bar b,c\bar c,gg)$ is less than $\approx 15\%$, and we expect
a drop in the upper limit to $m_h$ compared to $m_{H_{SM}}$ of only $\sim 1
\GeV$. Coupled with the very SM-like $h$ production (see Fig. 8) for $m_h\lsim
60\GeV$, and $R_h\gsim 0.85$, we find that the $m_{H_{SM}}>60\GeV$ limit also
applies to the $h$ Higgs of both the minimal and flipped $SU(5)$ models.
For a more detailed discussion of $h$ Higgs mass limits at LEPI in the two
models we consider here, see Ref. \LNPWZh.

\newsec{Discussion and Conclusion}
In this paper we have studied the most promising signals for supersymmetry at
LEPII in the context of two well motivated supergravity models: (i) the minimal
$SU(5)$ supergravity model including the stringent constraints from proton
stability and a not too young Universe, and (ii) a recently proposed
string-inspired no-scale flipped $SU(5)$ supergravity model. These signals
involve the neutralino/chargino, slepton, and  Higgs sectors. Because of the
study of {\it specific} models, we are led to modifications in the standard
assumptions regarding sparticle and Higgs decay. In the first sector we
computed the number of `mixed' (1-lepton + 2-jets + $\slash{p}$) events
occuring in the decay of pair-produced charginos ($\chi^\pm_1$) and found that
the predictions for both models should lead to detection (with ${\cal
L}=100\ipb$)
up to the kinematical limit ($m_{\chi^\pm_1}\lsim100\GeV$). Moreover, these
predictions do not overlap: the minimal model predictions being larger than the
flipped model ones. This result can be directly traced to a characteristically
light sneutrino spectrum in the flipped case ($m_{\tilde\nu}\simeq0.3m_{\tilde
g}$). This implies that if $m_{\chi^\pm_1}<100\GeV$ then LEPII should be able
to exclude at least one of the two models. In fact, in the minimal $SU(5)$
model $m_{\chi^\pm_1}<104\,(92)\GeV$ for $\mu>0\,(\mu<0)$, assuming $m_{\tilde
q,\tilde g}\lsim1\TeV$, while in the flipped case $m_{\tilde\chi^\pm_1}\lsim
285\GeV$ ($\mu>0,\mu<0$) and the mixed chargino signature can be suppressed.
Consequently, it is possible to explore nearly all of the allowed parameter
space for the minimal $SU(5)$ model but only $\lsim20\%$ of the flipped model.

We found significant chargino-dilepton event rates (per ${\cal L}=500\ipb$ for
$m_{\chi^\pm_1}>100\GeV$) in the flipped model, and a negligible signal in the
minimal model. The question of backgrounds to this process remains open.
The magnitude of the experimental efficiency cut for this dilepton signal is
not known at present. In the models we consider, the relations among the
neutralino an chargino masses $m_{\chi^\pm_1}\approx m_{\chi^0_2}\approx
2m_{\chi^0_1}$ (see Table I) imply that the $e^+e^-\to\chi^0_1\chi^0_2$ process
could in principle explore indirectly chargino masses up to $\sim130\GeV$.

The slepton sector could be kinematically accessible at LEPII only in the
flipped $SU(5)$ model. We studied $e^+e^-\to\tilde e^+_L\tilde e^-_L+\tilde
e^+_R\tilde e^-_R+\tilde e^\pm_L\tilde e^\mp_R$ and obtained significant
numbers of di-electron events which may allow exploration of the full
kinematical range with ${\cal L}=500\ipb$. Smuon and
stau production are suppressed but may be observable as well. Correlating the
slepton and chargino sectors we observed that slepton-dileptons could probe
indirectly chargino masses as high as $\sim150\GeV$, and thus $\sim50\%$ of
the allowed parameter space. This is especially
important for this (the flipped) model since a significant number of points in
parameter space for $m_{\chi^\pm_1}<100\GeV$ yield negligible mixed chargino
event signatures. We also discussed the impact of chargino-dileptons on the
slepton-dileptons and the possibilities for experimental discrimination of
these signals. For an analysis of the {\it total} dilepton signal from all
supersymmetric sources in these models see Ref. \LNPWZd.

In the Higgs sector we found that the cross section $e^+e^-\to Z^*h$ deviates
negligibly from the SM result in the minimal model, whereas it can be as much
as $1/3$ lower in the flipped model. Also, the usually neglected invisible mode
$h\to\chi^0_1\chi^0_1$ can erode the preferred $h\to b \bar b,c\bar c,gg$
($h\to b\bar b$) for $m_h\lsim80\GeV$ ($m_h\gsim80\GeV$) by as much as
$30\%/15\%$ ($40\%/40\%)$ in the minimal/flipped model. The $h\to gg$ mode is
usually below $\approx0.2$ although there are exceptional points in the minimal
model where it can be much larger, because of a very light $\tilde t_1$.

We have recently shown \LNPWZh\ that the current experimental lower bound on
the SM Higgs boson mass ($m_{H_{SM}}>60\GeV$) applies as well to both
supergravity models considered here, and therefore is more stringent than the
supposedly model-independent experimental lower bound $m_h>43\GeV$. In this
connection, we have found it useful to relate the results obtained in the
chargino sector (as shown in Fig. 3) with those obtained in the Higgs sector by
plotting the number of mixed events in chargino pair production  versus the
Higgs mass; this is shown in Fig. 13. With this plot it is straightforward to
determine which points of interest in the chargino sector become excluded by an
increasing lower bound on the Higgs mass. In particular, all points for
$m_h<60\GeV$ are actually experimentally excluded. At LEPII, if no Higgs
events are seen for $m_h\lsim80\GeV$, Fig. 13 shows that in the minimal model
$\gsim125\,(l+2j+\slash{p})$ events are expected. This would allow to
unambiguously test these models. In fact, the number of mixed chargino events
seen is predicted to be different in the two models for the same Higgs mass
limit.

We conclude that the charged slepton sector is a deeper probe
than the \hfill\break
chargino/neutralino or Higgs sectors of the flipped $SU(5)$ model at
LEPII, while the opposite is true for the minimal $SU(5)$ model where the
slepton sector is no probe at all. The interconnections among the various
sectors of the models should make them easily falsifiable, or, if verified
experimentally, hard to dismiss as coincidences thus providing firm evidence
for the underlying structure of these models.

\bigskip
\bigskip
\bigskip
\bigskip
\noindent{\it Acknowledgments}: We would like to thank M. Felcini, J.-F.
Grivaz, J. Hilgart, S. Katsanevas, and J. White for very helpful discussions.
J.L. would like to thank the CERN-Theory Division for its hospitality while
part of this work was being done. This work has been supported in part by DOE
grant DE-FG05-91-ER-40633. The work of J.L. has been supported by an SSC
Fellowship. The work of  D.V.N. has been supported in part by a grant from
Conoco Inc. The work of X. W. has been supported by a T-1 World-Laboratory
Scholarship. We would like to thank the HARC Supercomputer Center for the use
of their NEC SX-3 supercomputer and the Texas A\&M Supercomputer Center for the
use of their CRAY-YMP supercomputer.
\vfill\eject
\listrefs
\vfill\eject

\noindent{\bf Table I}: Comparison of the most important features describing
the minimal $SU(5)$ supergravity model and the no-scale flipped $SU(5)$
supergravity model.
\bigskip
\input tables
\thicksize=1.0pt
\leftjust
\parasize=2.6in
\begintable
Minimal $SU(5)$ supergravity model\| No-scale flipped $SU(5)$ supergravity
model\crthick
\para{Not easily string-derivable, no known\nextline examples}\|
\para{Easily string-derivable, several known\nextline examples}\nr
\para{Symmetry breaking to Standard Model due to vev of \r{24} and
independent of supersymmetry breaking}\|
\para{Symmetry breaking to Standard Model due to vevs of \r{10},\rb{10}
and tied to onset of supersymmetry breaking}\nr
\para{No simple mechanism for doublet-triplet splitting}\|
\para{Natural doublet-triplet splitting mechanism}\nr
\para{No-scale supergravity excluded}\|{No-scale supergravity by
construction}\nr
\para{$m_{\tilde q},m_{\tilde g}<1\TeV$ by ad-hoc choice:\nextline
naturalness}\|
\para{$m_{\tilde q},m_{\tilde g}<1\TeV$ by no-scale mechanism}\nr
\para{Parameters 5: $m_{1/2},m_0,A,\tan\beta,m_t$}\|
\para{Parameters 3: $m_{1/2},\tan\beta,m_t$}\nr
\para{Proton decay: $d=5$ large, strong\nextline constraints needed}\|
\para{Proton decay: $d=5$ very small}\nr
\para{Dark matter: $\Omega_\chi h^2_0\gg1$ for most of the parameter space,
strong constraints needed}\|
\para{Dark matter: $\Omega_\chi h^2_0\lsim0.25$, ok with cosmology and big
enough for dark matter problem}\nr
\par{$1\lsim\tan\beta\lsim3.5$, $m_t<180\GeV$, $\xi_0\gsim6$}\|
\par{$2\lsim\tan\beta\lsim32$, $m_t<190\GeV$, $\xi_0=0$}\nr
\par{$m_{\tilde g}\lsim400\GeV$}\|
\par{$m_{\tilde g}\lsim1\TeV$,
$m_{\tilde q}\approx m_{\tilde g}$}\nr
\par{$m_{\tilde q}>m_{\tilde l}>2m_{\tilde g}$}\|\par{$m_{\tilde l_L}\approx
m_{\tilde\nu}\approx0.3m_{\tilde g}\lsim300\GeV$}\nr
\par{}\|\par{$m_{\tilde l_R}\approx0.18 m_{\tilde g}\lsim200\GeV$}\nr
\par{$2m_{\chi^0_1}\sim m_{\chi^0_2}\sim m_{\chi^\pm_1}\sim
0.3 m_{\tilde g}\lsim100\GeV$}\|
\par{$2m_{\chi^0_1}\sim m_{\chi^0_2}\approx m_{\chi^\pm_1}\sim
0.3 m_{\tilde g}\lsim285\GeV$}\nr
\par{$m_{\chi^0_3}\sim m_{\chi^0_4}\sim m_{\chi^\pm_2}\sim\vert\mu\vert$}\|
\par{$m_{\chi^0_3}\sim m_{\chi^0_4}\sim m_{\chi^\pm_2}\sim\vert\mu\vert$}\nr
\par{$60\GeV<m_h\lsim100\GeV$}\|\par{$60\GeV<m_h\lsim135\GeV$}\cr
\par{}\|\par{Strict no-scale: $\tan\beta=\tan\beta(m_{\tilde g},m_t)$}\nr
\par{No analog}\|{$m_t\lsim135\GeV\Rightarrow\mu>0,m_h\lsim100\GeV$}\nr
\par{}\|{$m_t\gsim140\GeV\Rightarrow\mu<0,m_h\gsim100\GeV$}\cr
\par{Cosmic Baryon Asymmetry?}\|\par{Cosmic Baryon Asymmetry explained \ENO}

\endtable

\listfigs
\bye